\documentclass[preprint,showpacs,showkeys,prb,amsmath,amssymb]{revtex4}
\usepackage{graphicx}

\begin{document}

\title{Effect of interface alloying and band-alignment on the Auger recombination of heteronanocrystals}

\author{J.I. Climente}
\author{J.L. Movilla}
\author{J. Planelles}
\email{josep.planelles@uji.es}
\affiliation{Departament de Qu\'{\i}mica F\'{\i}sica i Anal\'{\i}tica,
Universitat Jaume I, Castell\'o de la Plana, Spain}
\date{\today}

\keywords{Auger recombination, quantum dot blinking, core/shell nanocrystal, type-II structure}

\begin{abstract}
We report a numerical study of the effect of interface alloying and band-alignment
on the Auger recombination processes of core/shell nanocrystals. 
Smooth interfaces are found to suppress Auger recombination, the strength of the 
suppression being very sensitive to the core size.
The use of type-II structures constitutes an additional source of suppression,
especially when the shell confines electrons rather than holes.
We show that ``magic'' sizes leading to negligible Auger recombination 
[Cragg and Efros, Nano Letters {\bf 10} (2010) 313]
should be easier to realize experimentally in nanocrystals with extended
interface alloying and wide band gap.
\end{abstract}

\pacs{73.21.La,78.67.Hc,79.20.Fv}

\maketitle

\section{Introduction}

\begin{figure}[h]
\includegraphics[width=0.45\textwidth]{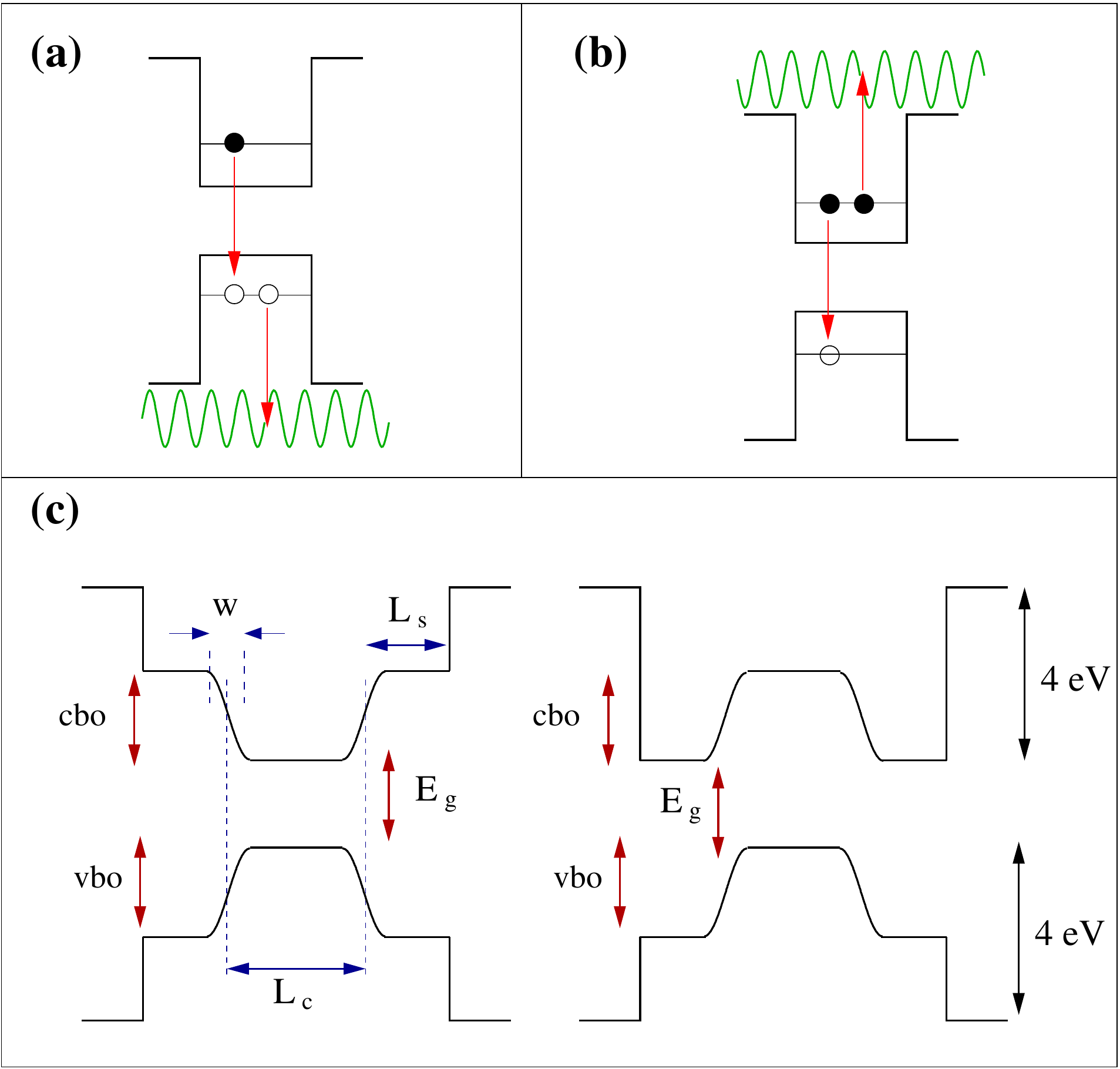}
\caption{Schematic of the Auger process under study, 
assuming the NC contains a positive (a) and negative (b) trion.
(c) Confinement potential profile used for type-I (left) 
and type-II NCs (right).}
\label{Fig1}
\end{figure}

There is current interest in minimizing Auger recombination (AR) processes
in semiconductor nanocrystals (NCs). This is because such processes are
considered to be responsible for the undesired blinking of NCs, which 
hinders their use in optical applications.\cite{EfrosNM}
When an electron-hole pair is generated in a NC by absorption of light,
there is a possibility that one of the two carriers becomes trapped at the surface. 
When the next electron-hole pair is created, it combines with the remaining carrier.
If the remaining carrier is a hole, the NC now contains a positive trion 
(as in Fig.~\ref{Fig1}a). If it is an electron, it contains a negative trion
(as in Fig.~\ref{Fig1}b).
At this point there is a competition between radiative electron-hole recombination
 and non-radiative Auger recombination, by which the energy of the recombining
electron-hole pair is transferred to the extra carrier -as shown in Figs.~\ref{Fig1}a and \ref{Fig1}b-. 
The excess carrier then moves into a highly excited (typically unbound) state and rapidly loses 
kinetic energy to heat.
For optical emission to be free from blinking, 
the AR process must be slower than the radiative recombination one.

Two promising techniques in slowing down AR processes have been developed in
the last years: (i) the use of hetero-NCs with radially graded composition,\cite{WangNAT}
and (ii) the use of hetero-NCs with quasi-type-II or type-II band alignment.\cite{OronPRB,
MahlerNM,SpinicelliPRL,OsovskyPRL,SantamariaNL,SantamariaNL2,VelaJBP}
Theoretical understanding of the AR suppression in (i) was provided by
Cragg and Efros.\cite{CraggNL} They showed that the smooth confinement potential 
resulting from the graded composition removes high-frequency Fourier components
of the electron and hole wave functions, which in turn reduces AR rates.
Theoretical understanding of the AR suppression in (ii) was provided by us
in Ref.~\onlinecite{ClimenteSmall}, where we showed that it is originated in
the removal of high-frequency components of the carrier that penetrates into 
the shell.

In this work, we extend Refs.~\onlinecite{CraggNL,ClimenteSmall} by 
performing a systematic study of the influence of smooth confinement 
potentials and band alignment on the AR rates of hetero-NCs.
We consider not only AR processes involving excess holes (Fig.~\ref{Fig1}a),
but also excess electrons (Fig.~\ref{Fig1}b).
We find that, as noted in Ref.~\onlinecite{CraggNL}, the softer
the confinement potential the slower the AR. 
However, the magnitude of the AR suppression strongly depends on 
the core sizes, with order-of-magnitude differences.
A similar behavior is observed when varying the core-shell band-offset.
In general, moving from type-I to type-II hetero-NC translates into
slower AR rates, but the actual value is very sensitive to the NC dimensions.
The decrease of the AR rate is particularly pronounced when the carrier confined
in the shell is the electron instead of the hole. 
Last, we show that the narrow ranges of (``magic'') NC sizes leading to almost 
complete suppression of AR reported in Refs.~\onlinecite{CraggNL,ChepicJL}  
become wider in NCs with smooth potential or wide band gap.

\section{Theory}

In our calculations, we describe electrons and holes with a one-dimensional, 
two-band Kane Hamiltonian\cite{ForemanPRB}:

\begin{equation}
H = 
\left( 
\begin{array}{cc}
-\frac{p^2}{2\,m_e} + V_c(x) + E_g/2 	&   K\,p \\
K\,p					& \frac{p^2}{2\,m_h} - V_v(x) - E_g/2 
\end{array}
\right).
\label{H_kane}
\end{equation}

\noindent where $p=-i\hbar\,\partial / \partial x$ is the momentum operator,
$m_{e(h)}$ is the effective mass of the electron (hole) disregarding
the influence of the valence (conduction) band, hereafter VB (CB). 
$E_g$ is the energy gap as defined in Fig.~\ref{Fig1}c, $K$ is the Kane 
matrix element and $V_{c(v)}(x)$ the CB (VB) confinement potential. 
The heigth of $V_{c(v)}(x)$ is given by the core/shell band-offset and the shape
is depicted in Fig.~\ref{Fig1}c. Note that, at the core-shell interface, $V_{c(v)}(x)$
has a cosine-like profile of width $w$, which accounts for interfase diffusion.
Varying $w$ we can monitor the evolution from abrupt interfaces ($w=1$ \AA) to 
alloyed interfaces spread over several monolayers ($w$ tens of \AA),
accounting for either spontaneous interfase diffusion\cite{SantamariaNL2} or 
intentionally graded composition.\cite{WangNAT}
Equation (\ref{H_kane}) is integrated numerically using a finite differences scheme.
The resulting electron and hole states have a mixture of CB and VB components:

\begin{equation}
\psi_{i}= \chi_{i}(x)\,u_c(x) + \eta_{i}(x)\,u_v(x), 
\end{equation}

\noindent with $i\,=\,e,h$.  
Here $\chi$ and $\eta$ ($u_c$ and $u_v$) are the envelope (periodic) parts of the 
Bloch wave function corresponding to the CB and VB, respectively. 
For electrons, the CB component $\chi$ is by far dominant.
The opposite holds for holes.

The AR rate is calculated using Fermi's golden rule:

\begin{equation}
\tau_A^{-1} = \frac{2\pi}{\hbar} \, \left| \langle i | V | f \rangle \right|^2 \, \rho(E_f)\,\delta(E_i - E_f).
\label{fermi}
\end{equation}

Here, $V(r)=1/ (\epsilon\,(r+\Delta))$ is the Coulomb potential, with $\epsilon$ the dielectric constant, 
$r=|x_1-x_2|$ and $\Delta$ a parameter introduced to avoid the Coulomb singularity.
$E_i$ and $E_f$ are the energies of the initial
and final states, respectively. 

To proceed further, we consider the NC are in the strong confinement regime. 
Let us first assume the most relevant AR process is that illustrated in Fig.~\ref{Fig1}a. 
This seems to be the case at least in CdSe/ZnS and CdZnSe/ZnSe NCs\cite{WangNAT,HeyesPRB}.
The initial state, $|i\rangle$, is then defined by 

\begin{equation}
|i\rangle = \psi_h^0 (x_1)\,  \psi_h^0 (x_2)\,\frac{1}{\sqrt{2}}\,\left( \alpha (1)\, \beta(2)  - \beta(1)\,\alpha(2) \right),
\end{equation}

\noindent with $\alpha$ ($\beta$) standing for spin up (down) projections.
The final state, $|f\rangle$, is: 

\begin{equation}
|f\rangle = \frac{1}{\sqrt{2}} ( \psi_e^0 (x_1)^*\,  \psi_h^{ex} (x_2) + \psi_h^{ex} (x_1) \, \psi_e^0 (x_2)^* )
\,\frac{1}{\sqrt{2}}\,\left( \alpha (1)\, \beta(2)  - \beta(1)\,\alpha(2) \right).
\end{equation}

\noindent The excited, unbound hole state is taken as $\psi_h^{ex} (x) = \sqrt{\frac{1}{L}} \,e^{i\,k\,x}\,u_v$,
i.e. a free electron plane wave with negligible component in the CB. 
$L$ is the length of the computational box. 
The momentum $k$ is determined by energy conservation,
$k=\sqrt{2\,m_0\,(E_e^0 + 2\,E_h^0 + E_g)}/\hbar$, where $m_0$ is the free electron mass,
while $E_e^0$ and $E_h^0$ are the single-particle electron and hole energies 
defined with respect to the CB and VB edges, respectively.
$\rho (E_f)$ is the density of states of $|f\rangle$, which can be
approximated by that of the hole in the continuum (computational box),
$\rho (E_f) \approx L/\hbar \pi\;\sqrt{m_0/2\,E_h^{ex}}$,
 where $E_h^{ex}=(k^2/2m_0 - \mbox{vbo}) $ is the single-particle 
energy of the plane wave substracting the VB offset.
Note that $L$ in $\rho(E_f)$ cancels the $L$ factor arising from the plane wave 
normalization constants in $|\langle i|V|f\rangle|^2$.

Because $\psi_h^{ex} (x)$ has no CB component,
the Coulomb matrix element in Eq.~(\ref{fermi}) is given by:

\begin{equation}
\langle i | V | f \rangle =  \sqrt{2} \left(
\langle \chi_h\,\eta_h|\,V\,|\chi_e\,e^{i k x_2} \rangle 
+
\langle \eta_h\,\eta_h|\,V\,|\eta_e\,e^{i k x_2} \rangle
\right).
\label{integrals}
\end{equation}

For computational performance and physical insight, 
it is convenient to rewrite the above integrals in the
Fourier space. For example,

\begin{equation}
\langle \chi_h\,\eta_h|\,V\,|\chi_e\,e^{i k x_2} \rangle = 
\frac{1}{2\pi}\,\int_{-\infty}^{\infty} \,dq \,{\hat V}(q)\,{\hat \chi}(q)\,{\hat \eta}(q),
\label{fourier}
\end{equation}

\noindent where ${\hat V}(q)=(e^{-i q \Delta}\,E_1(-i q \Delta) + e^{i q \Delta}\,E_1 (i q \Delta))/\epsilon$
is the Fourier transform of $V(r)$, with $E_1(x)=\int_x^{\infty} e^{-t}/t\,dt$ 
the exponential integral.\cite{Abramowitz_book}
${\hat \chi}(q) = \int\,dx_1\,\chi_h(x_1)^*\,\chi_e(x_1)\,e^{i q x_1}$,
 is the Fourier transform of the CB components and
${\hat \eta}(q) = \int\,dx_2\,\eta_h(x_2)^*\,\,e^{i (k - q) x_2}$
that of the VB components. 

An analogous development can be carried out for the case of Fig.~\ref{Fig1}b,
where the relevant AR process is that involving a negative trion. In such a case:

\begin{equation}
\langle i | V | f \rangle =  \sqrt{2} \left(
\langle \eta_e\,\chi_e|\,V\,|\eta_h\,e^{i k x_2} \rangle 
+
\langle \chi_e\,\chi_e|\,V\,|\chi_h\,e^{i k x_2} \rangle
\right).
\label{integrals2}
\end{equation}

For the simulations, we take material parameters close to those of typical
II-VI NCs: $m_e=0.15\,m_0$, $m_h=0.6 \, m_0$, $E_g=1.5$ eV, $K=(21\,eV/2 m_0)^{1/2}$, 
$\epsilon=6$ and $\Delta=10^{-2}$ \AA. 
The CB (VB) offset at the shell is cbo$=0.5$ eV (vbo$=0.5$ eV), unless otherwise stated.
A high potential barrier, $4\,eV$, is set at the external medium. 

\section{Results and discussion}

\subsection{Effect of interface alloying}

\begin{figure}[h]
\includegraphics[width=0.45\textwidth]{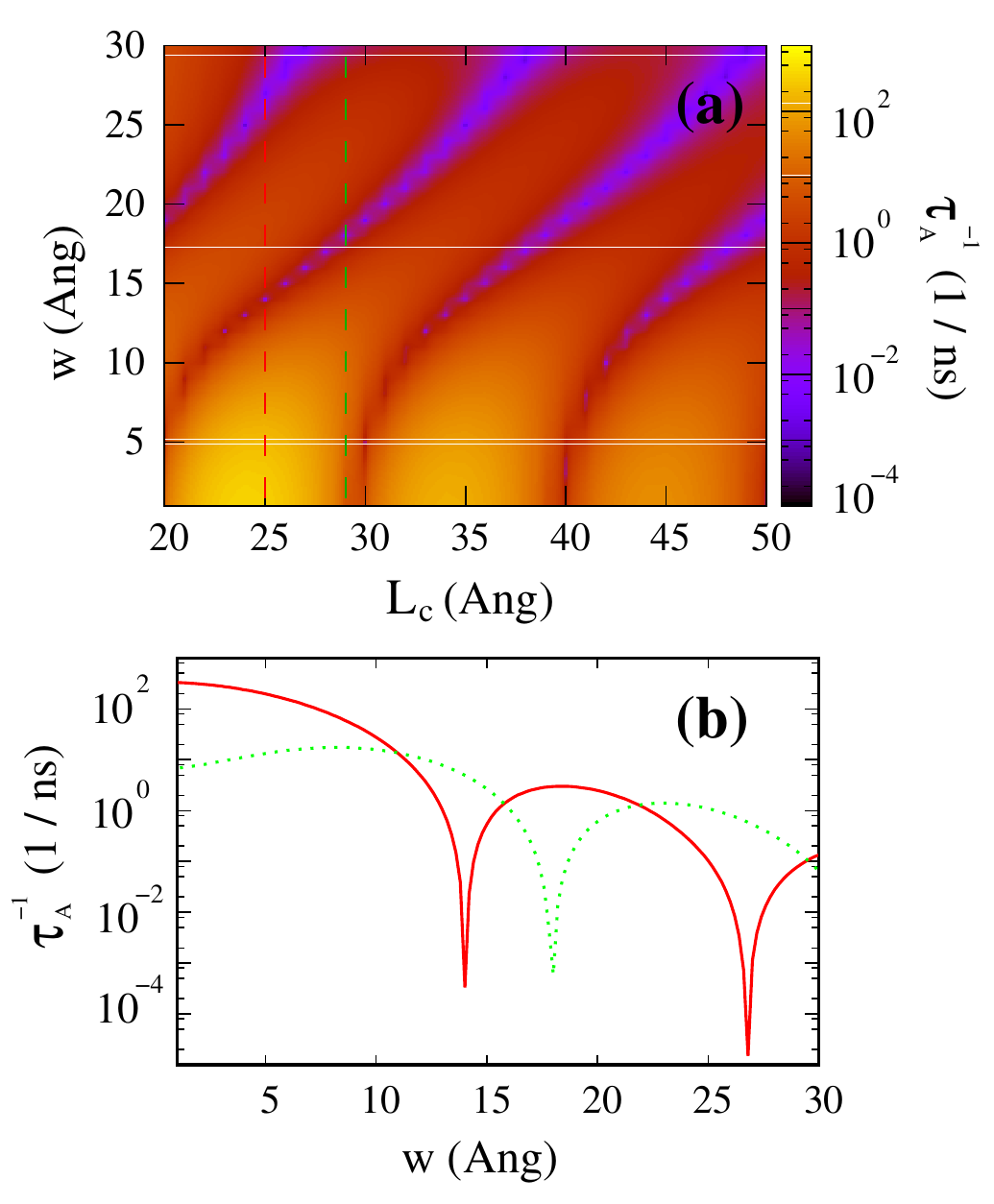}
\caption{(a) Contour plot of the AR rate as a function of core size and interfase thickness in type-I NCs. 
(b) Vertical cross-sections at $L_c=25$ \AA\ (solid line) and $L_c=29$ \AA\ (dotted line).
The AR process considered is that involving an excess hole.}
\label{Fig2}
\end{figure}

We start by studying a type-I core/shell NC, where electrons and holes are well confined in the core,
surrounded by a $L_s=20$ \AA\ shell acting as a barrier.  
We consider AR involves an excess hole (Fig.~\ref{Fig1}a scheme) and
model the influence of diffusion around the interface by shaping the confining potential
from abrupt ($w=1$ \AA) to smooth (diffusion spreading over $w=30$ \AA) for different core sizes.
The resulting AR rates are plotted in Fig.~\ref{Fig2}a.  As expected from simple ``volume''
scaling\cite{KlimovSCI,CraggNL}, in general the bigger the core the slower the AR rate.
In addition, for a given $w$, one can see that increasing the core size leads to
periodic valleys where AR is strongly suppressed (e.g. at $L_c \approx 20,\,30,\,40$ \AA\ for $w=1$ \AA).
These ``magic sizes'' were first predicted by Efros and co-workers.\cite{ChepicJL}
They are originated in destructive quantum interferences between the initial and final 
states of the Auger scattering, and should be detectable at low temperature and pumping power.\cite{CraggNL}
We now pay attention to the influence of interface smoothness.
In general, softening the interface potential reduces the AR by up to two orders of magnitude.\cite{CraggNL,ChepicJL}
However this trend is neither universal nor monotonic because, as can be seen in Fig.~\ref{Fig2}a,
softening the potential also changes the position of the suppression valleys.
This point is exemplified in Fig.~\ref{Fig2}b, which illustrates the vertical cross-sections
highlighted in Fig.~\ref{Fig2}a with dashed lines. For some core sizes ($L_c=25$ \AA)
there is a strong decrease of $\tau_A^{-1}$ with $w$. For others, which are initially in a valley of
AR ($L_c=29$ \AA), softening the potential provides no benefit up to large values of $w$ ($w>12$ \AA).
The shift in the position of the suppression valleys with increasing $w$ is due to the decrease of the 
effective NC size as the shell material diffuses into the core.
Interestingly, while the valleys are very narrow for $w=1$ \AA\ ($\Delta L_c \sim 1$ \AA),  
they become wider for $w=30$ \AA\ ($\Delta L_c \sim 4$ \AA). 
This should facilitate the experimental observation of these minima, since the most
precise growth techniques rely on monolayer deposition ($\sim 4$ \AA\ each monolayer
in typical II-VI NCs).

\begin{figure}[h]
\includegraphics[width=0.45\textwidth]{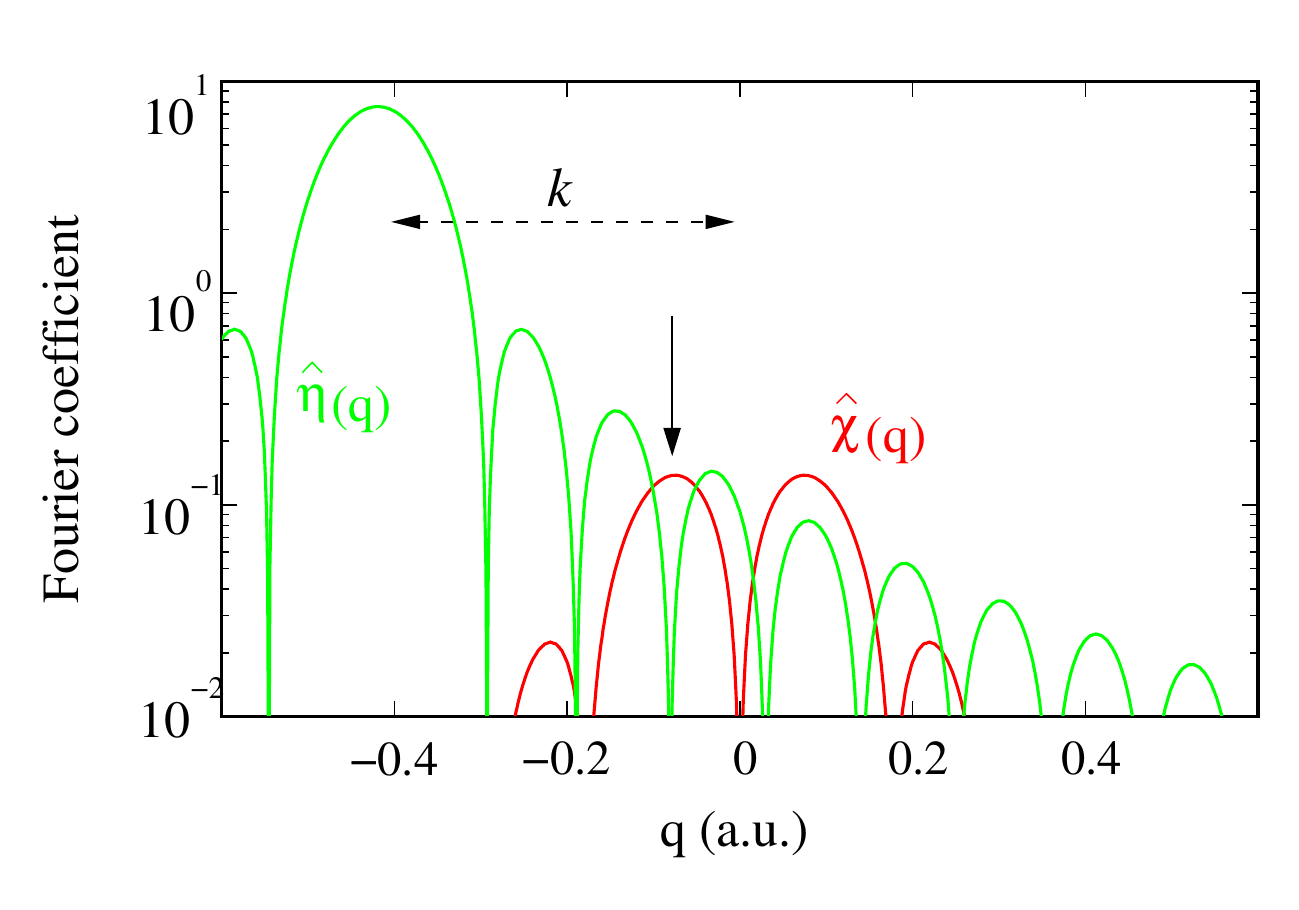}
\caption{Fourier transforms of the CB and VB terms in the Auger matrix element of Eq.~(\ref{fourier}).
The calculation corresponds to an AR minimum of Fig.~\ref{Fig2}a, namely $L_c=29$ \AA\ and $w=1$ \AA.}
\label{Fig3}
\end{figure}

Next we provide some insight into the conditions favoring the appearance of magic sizes.
In the reciprocal space, the minima occur when there is a near cancellation of Eq.~\ref{fourier}.
Fig.~\ref{Fig3} shows a typical Fourier spectrum of ${\hat \chi}(q)$ and ${\hat \eta}(q)$
at a magic size.\cite{Vnocal}
 Note that, while ${\hat \chi}(q)$ is centered at $q=0$, 
${\hat \eta}(q)$ is centered at $q=k$. If the $k$ value is such that one of the periodic zeros 
of ${\hat \eta}(q)$ coincides with the nearby maximum of ${\hat \chi}(q)$, as in the figure 
--see vertical arrow--, ${\hat \eta}(q)$ and ${\hat \chi}(q)$ are in anti-phase and the AR minimum is obtained.
The NC sizes which lead to this situation can be approximated with a simple model.
Let the potential in Eq.~\ref{H_kane} be that of a square well with infinite walls
and width $L_c$. If we include interband coupling by perturbation theory on the basis
of the two lowest CB and VB states, the ground state wave functions are:

\begin{eqnarray}
\psi_e^0 &=& a_e\,\cos{\left( \frac{\pi\,x}{L_c} \right)}\, u_c + 
             b_e\,\sin{\left( \frac{2\,\pi\,x}{L_c} \right)}\, u_v\\
\psi_h^0 &=& a_h\,\sin{\left( \frac{2\,\pi\,x}{L_c} \right)}\, u_c + 
             b_h\,\cos{\left( \frac{\pi\,x}{L_c} \right)}\, u_v.
\end{eqnarray}

\noindent where $a$ and $b$ are constants. 
The Fourier transform of the VB components, ${\hat \eta}(q)$, 
 has zeros at $q_0=k\pm \frac{\pi}{L_c}\,(2n+1)$, with $n=1,2,3\ldots$.
On the other hand, the Fourier transform of the CB components, ${\hat \chi}(q)$, 
 has the first local maximum at  $q_{M}\,L_c \approx -2\pi$.
We can then force $q_M=q_0$, obtaining $k = -\pi\,( 2n+3 )/L_c.$
Since $k=-\sqrt{2\,m_0\,(E_e^0 + 2\,E_h^0 + E_g)}/\hbar$,
using particle-in-the-box energies we get:

\begin{equation}
\sqrt{m_0\,\left( \frac{\pi^2}{m_e L_c^2} + 2\, \frac{\pi^2}{m_h L_c^2} + \frac{2\,E_g}{\hbar^2} \right)}= \frac{\pi}{L_c}\,( 2n+3 ).
\label{supp_IyqII}
\end{equation}

\noindent Isolating $L_c$ in the above expression leads to:

\begin{equation}
L_c = \frac{\hbar \pi \, \sqrt{(2n+3)^2 - ( m_0/m_e + 2 m_0/m_h )}}{\sqrt{2\,E_g\,m_0}},
\label{supp_I}
\end{equation}

For the system in \ref{Fig2}a, Eq.~(\ref{supp_I}) predicts ``magic'' NC sizes
at $L=31,\,42$ and $53$ \AA, in close agreement with the numerical results at $w=1$ \AA.
The expression also reveals that wide gap materials are more prone to display AR minima.

\begin{figure}[h]
\includegraphics[width=0.45\textwidth]{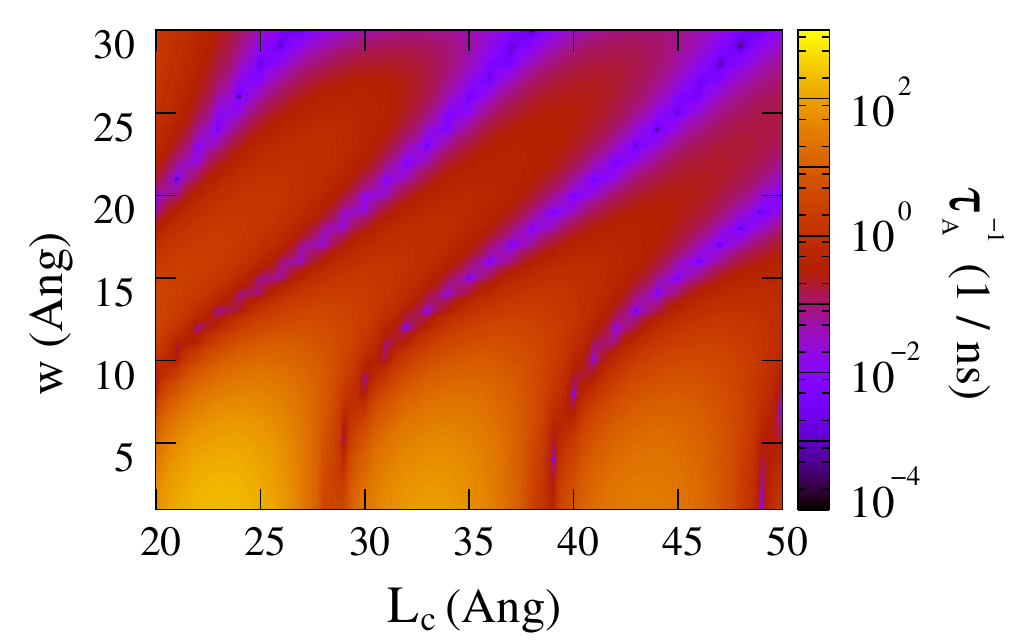}
\caption{Same as Fig.~\ref{Fig2}(a) but for an excess-electron process.}
\label{Fig4}
\end{figure}

So far we have considered AR processes involving an excess hole.
Very similar results are obtained if the process involves an excess electron
instead. To illustrate this point, in Fig.~\ref{Fig4} we reproduce Fig.~\ref{Fig2}a 
but now considering the negative trion process. In general, the AR are a few-times
slower, but the trends are the same and even the position of the AR minima are
nearly the same. This implies that magic sizes can suppress AR for either of the
two process.

\subsection{Effect of band alignment}

\begin{figure}[h]
\includegraphics[width=0.45\textwidth]{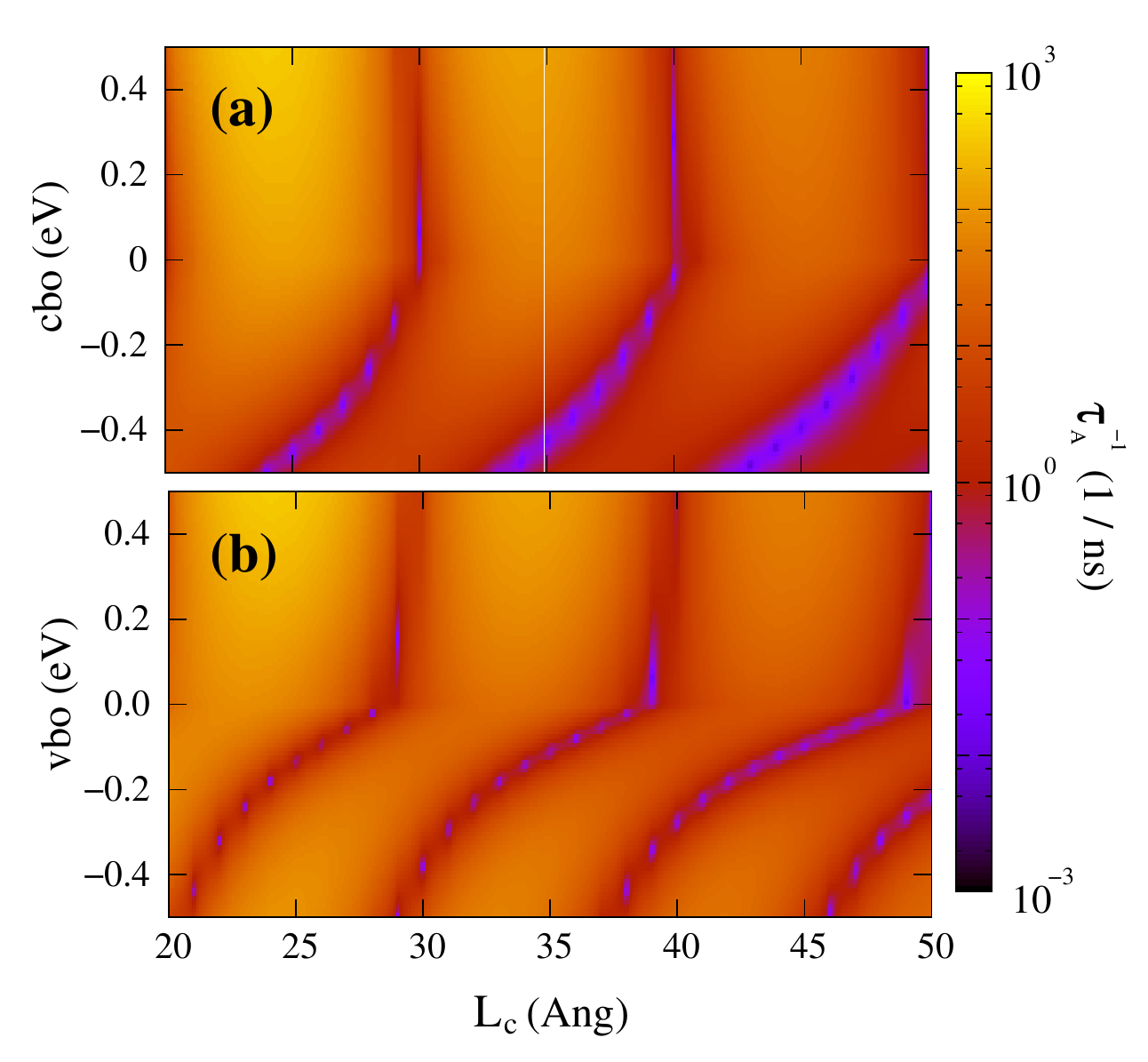}
\caption{Contour plot of the AR rate as a function of core size and 
core/shell band-offset.  (a) variable cbo; (b) variable vbo.
The shell thickness is $L_s=20$ \AA\ and $w=1$ \AA.}
\label{Fig5}
\end{figure}

In this section we study how the core/shell potential height influences the AR rates. 
CB and VB offsets are initially set at $0.5$ eV, then we lower the cbo --Fig.~\ref{Fig5}a-- 
or the vbo --Fig.~\ref{Fig5}b--.
One can see that lowering the offsets from $0.5$ eV to $0.0$ eV has a rather weak 
effect on $\tau_A^{-1}$. However, moving to negative offsets (i.e. switching 
from type-I to type-II band-alignment) 
brings about important modifications. Namely: (i) the position of the AR suppression valleys
is shifted. This is due to the changes in the effective NC size seen by electrons and holes.
(ii) The width of the suppression valleys increases, especially for negative cbo (Fig.~\ref{Fig5}a).
(iii) Lowering cbo reduces $\tau_A^{-1}$ by orders of magnitude. This is in agreement
with the experiments of Oron et al.\cite{OronPRB}, where using CdTe/CdSe NC --with the
hole confined in the core and the electron in the shell-- lead to a strong decrease of the AR rate.
Surprisingly, lowering vbo instead (Fig.~\ref{Fig5}b) barely reduces $\tau_A^{-1}$.
In other words, using CdSe/CdTe NCs would provide much less benefit.

\begin{figure}[h]
\includegraphics[width=0.45\textwidth]{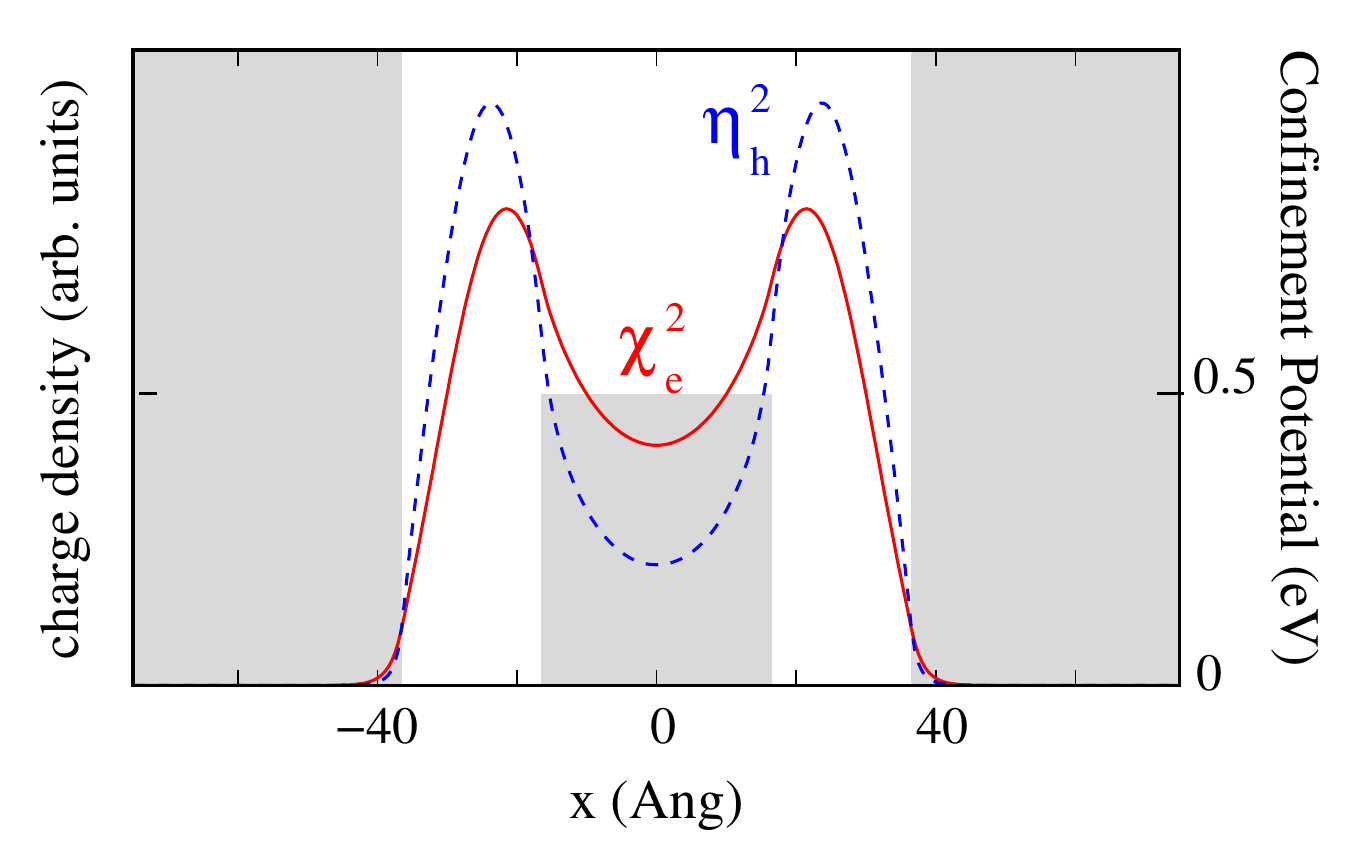}
\caption{Density charge of the dominant electron and hole wave function
components in a type-II NC where the carrier is confined in the shell.
The gray area represents the confinement barrier.
$L_c=33$ \AA\ and $L_s=20$ \AA.}
\label{Fig6}
\end{figure}

The origin of the asymmetric behavior of electrons and holes in Fig.~\ref{Fig5}
can be ascribed to the different effective masses. To clarify this point,
in Fig.~\ref{Fig6} we compare the density charge of the dominant electron 
and hole wave function components in type-II NCs. Because of the lighter
effective mass, the electron ($\chi_e^2$) displays stronger tunneling across the core.
The resulting wave function is smoother, which translates into less 
high-frequency Fourier components than the hole, and hence slower AR.

We close by noting that in three-dimensional systems, quantitative
differences will probably arise from our estimates. Yet, the 
different role of cbo and vbo should persist, 
 as the effective masses of the two carriers are still different.  
Also, we have checked that the weak effect of vbo on the AR rate 
holds for processes involving an excess electron.

\section{Conclusions}

Using a two-band Kane Hamiltonian, we have studied how the interface potential 
shape and height influence the AR rates of heteronanocrystals. 
As previously noted\cite{CraggNL,ChepicJL}, a smooth confinement potential 
--which may result from either spontaneous interfase diffusion or intentionally 
graded composition--, reduces the AR rate. However, the magnitude of this
reduction strongly depends on the core size. This is due to the dependence
of the ``magic'' sizes suppressing AR on the interface thickness. 
Indeed, the range of ``magic'' sizes increases with the interface thickness,
which should facilitate their experimental detection.
Switching from type-I to type-II band-alignment further reduces the AR.
For moderate band-offsets (fraction of eV), this effect is more pronounced 
when the shell hosts the electron (instead of the hole). This is because
the stronger tunneling of electrons enables the formation of smooth wave functions.
These results are valid for AR processes involving either an excess electron 
or an excess hole.

\begin{acknowledgments}

Support from MCINN projects CTQ2008-03344 and CTQ2011-27324, 
UJI-Bancaixa project P1-1A2009-03 and the Ramon y Cajal program (JIC) is acknowledged.

\end{acknowledgments}


\begin{thebibliography}{20}
\bibitem{Vnocal} ${\hat V}(q)$ is a slowly-varying function with secondary influence
on the integral value.\cite{ClimenteSmall}
\end{thebibliography}
\end{document}